\magnification=1200
\vsize = 8.6 true in
\leftskip=1.0 cm
\rightskip 20 truept
\baselineskip=20 pt
$$\ \ $$

\bigskip
\bigskip
\bigskip
 
\centerline {\bf Electromagnetic Force as consequence of the Geometry of	Minkowskian Spacetime}

\bigskip
\centerline {J. Buitrago}

\bigskip
\centerline  {Departament of Astrophysics of the University of La Laguna}
\centerline {Avda. Astrofisico Francisco Sanchez s/n.}
\centerline {38207 La Laguna (Tenerife) Spain}
\bigskip
\centerline {email: jgb@iac.es}
\centerline {******}
\bigskip

\bigskip
\centerline {PACS numbers: 00., 03., 03.50.-z, 03.50.De}
 
\bigskip

\vfill
\eject

\bigskip
{\noindent \bf Abstract}

By describing the dynamical evolution of a test charged particle in the
presence of
 an  electromagnetic field as a succession of infinitesimal
Lorentz boosts and rotations it is possible to obtain 
the Lorentz Force of Electrodynamics.
A consequence of this derivation at the classical level is
that, given the existence of electric and magnetic fields,
 the form of the electromagnetic force acting
on the particle can be regarded as  
arising  from the geometry 
of Minkowskian spacetime.
\vfill
\eject
\bigskip

\noindent {\bf 1. Introduction}

As is well known, the Lorentz Force acting on a charged test particle
under the action of an 
electromagnetic field cannot be obtained as a consequence
of Maxwell Equations, but must be postulated in an independent manner. In
some text books, heuristic arguments about the form of the force are given,  
such as being the simplest
 four-force compatible with special relativity and linear in the
components of the four-velocity (Barut 1964, Rindler 1982),
while in other undergraduate texts  it
is attempted to derive the existence
 of  magnetic fields (and even the specific form of the magnetic force) 
from Coulomb's law of electrostatics and the
Lorentz transformations of special 
relativity (Lorrain and Corson 1970). Another
approach to the problem is to make Newton's second law	applied to a particle
in an electrostatic field consistent with relativistic invariance (Kobe, 1986).

The problem 
of the relation between the 
dynamical behaviour of charged bodies  in
electromagnetic fields
and special relativity has been the subject of some
controversy  which we consider out of the scope 
of this work. 
A usual procedure which can be
found in many text books is to	consider the equation of motion of a
particle in an inertial frame, in which there is 
only an electric field, and
then transform to another inertial frame. As a matter of fact,
from this approach it is possible
to obtain some partial features of electric and magnetic forces (see 
the remarks of Jackson, 1975) but not
to obtain the electromagnetic force which, as we shall attempt to show in this
article, is not related to the usual finite Lorentz transformations but to
infinitesimal boosts and rotations in spacetime.

The motivation for this article is to present a  derivation of the
covariant equations of motion of a charged particle in an
electromagnetic field that suggests a new
interpretation of the Lorentz Force  as an  unavoidable
consequence of the geometric structure of Minkowskian spacetime
and also gives new insight into the nature of electric and magnetic 
fields.

Central to our discussion will be a somewhat detailed examination of how
the evolution of the {\it four velocity vector} of a test particle can
be described by means of infinitesimal boosts and rotations and how
these transformations are 
seen from the standpoint of two different inertial frames
of reference,
this is done in sections 2 and 3. In section 4 we shall relate
the equations of motion obtained from a
description in terms
of parallel axis boosts to the motion of a charged test particle under the
action of an electric field. However, the equation of motion, in this case,
are not  susceptible
of being written in a
covariant tensor form. We shall see that a more general 
description based in the
combination of boosts and rotations leads in a natural way to the covariant
form of the electromagnetic force.

\bigskip
\bigskip

{\noindent \bf 2. Infinitesimal boosts and rotations}
\bigskip

Suppose we have a particle of mass $m$ following an accelerated, not
neccessarily rectilinear,
trajectory relative to a certain inertial frame $S$. At a generic
proper time $\tau$ the four velocity in frame $S$ is $\bar u(\tau)$ and
after a small lapse of proper time $\delta\tau$ the four velocity has
evolved to
$\bar u(\tau+\delta\tau)$ relative to the same frame. By the
principle of relativity it is always possible to choose another inertial frame
$S^*$ such that at proper time $\tau$ the four velocity of the particle is

$$\bar u^*(\tau)= \bar u(\tau+\delta\tau).\eqno(1)$$
Since $S$ and $S^*$ are inertial frames we know that there is a 
Lorentz transformation, a spatial rotation
or a combination of both that relate  $\bar u(\tau)$
and $\bar u^*(\tau)$. Furthermore, from the assumed smallness of $\delta\tau$
the transformation need only be infinitesimal. If $A(\tau)$ is such a
transformation we can, accordingly, write

$$
\bar u^*(\tau)=A(\tau)\bar u(\tau), \eqno(2)
$$
and therefore
$$
\bar u(\tau+\delta\tau)=A(\tau)\bar u(\tau). \eqno(3)
$$

\bigskip
Departing here from the conventional use of Lorentz boost as transformations
relating geometric objects in two {\it different} inertial frames, we shall,
henceforth,
regard	$A(\tau)$ as a linear operator relating $\bar
u(\tau+\delta\tau)$ and $\bar u(\tau)$ {\it in the  same} frame of
reference.

In the case that the change in the particle's four velocity can be
described by an infinitesimal Lorentz boost alone, the explicit form of
$A(\tau)$ is a $4\times 4$ matrix

$$
	A(\tau) =e^{\delta\vec v.\vec K}\simeq I+\delta\vec v.\vec K, 
	\eqno(4)
	$$
 $\delta\vec v$ being the change in the
velocity of the body during the lapse of
time $\delta\tau$ as measured in $S$ 
(we are using units in which $c=1$ and signature of the special 
relativistic metric tensor $+ - - -$), $I$ the unit matrix,
 and the components of
$\vec K$ the generators of Lorentz boosts (see Jackson, 1975)

\bigskip

$$
K_1=\left( \matrix{0	    & 1      & 0      & 0   \cr
			1	& 0	 & 0	  & 0	\cr
			0	& 0	 & 0	  & 0	\cr
			0	& 0	 & 0	  & 0	\cr}
			\right) ,
K_2=\left( \matrix{0	    & 0     &  1      & 0   \cr
			0      & 0	& 0	 & 0   \cr
			 1	 & 0	  & 0	   & 0	 \cr
			0	& 0	 & 0	  & 0	\cr}
			\right) ,
K_3=\left( \matrix{0	    & 0     & 0      &	1   \cr
			0      & 0	& 0	 & 0   \cr
			0	& 0	 & 0	  & 0	\cr
			1	& 0	 & 0	  & 0	\cr}
			\right). \eqno(5)$$

\bigskip

When the accelerated  motion follows a curved trajectory one
should, in general, include spatial rotations as well as boosts
which means that we need all the six generators of the Lorentz Group
(as infinitesimal boosts and rotations commute, see
equation 6 below, the order in which the transformations
are carried out is irrelevant).
Equation
(4) should be replaced
by

$$
	A=e^{-(\delta\vec\phi.\vec S-\delta\vec v.\vec K)}, \eqno(6)
	$$
where the components of $\vec S$ are the generators
of spatial rotations, explicitely:
\bigskip

$$S_1=\left( \matrix{0	      & 0     & 0      & 0   \cr
			0      & 0	& 0	 & 0   \cr
			0	& 0	 & 0	  & -1	 \cr
			0	& 0	 & 1	  & 0	\cr}
			\right) , 
S_2=\left( \matrix{0	    & 0     & 0      & 0   \cr
			0      & 0	& 0	 & 1   \cr
			0	& 0	 & 0	  & 0	\cr
			0	& -1	  & 0	   & 0	 \cr}
			\right) ,
S_3=\left( \matrix{0	    & 0     & 0      & 0   \cr
			0      & 0	& -1	  & 0	\cr
			0	& 1	 & 0	  & 0	\cr
			0	& 0	 & 0	  & 0	\cr}
			\right). \eqno(7) $$
\bigskip

\bigskip
{\noindent \bf 3. Transformation properties of the parameters of
infinitesimal boosts and rotations}
\bigskip

Consider two inertial systems $S$ and $S'$ with
coordinates $(x_0,x_1,x_2,x_3)$ and $(x'_0,x'_1,x'_2,x'_3)$ respectively.
The coordinate axes in the two frames are parallel and oriented so that the
system $S'$ is moving in the positive $x_1$ direction with velocity $v$, as
viewed from $S$.
 In $S'$ we have a
test particle  moving, at proper time $\tau$, in
the direction of the $x'_2$ axis with velocity
$v'$. Its four velocity  being

$$
\bar u'(\tau)=\pmatrix{{u'}^0 \cr
		0\cr
		{u'}^2 \cr
		0\cr}, \eqno(8)
$$
with $u'^2=\gamma' v'$, ($\gamma'=(1-v'^2)^{-1/2}$).

At proper time $\tau$, and always in $S'$, the particle accelerates in the $x'_2$
direction  during a lapse of proper time $\delta\tau$
thereby suffering a change in its velocity
$$
\delta\vec v'=\pmatrix{0\cr
			\delta v'_2\cr
			0\cr}. \eqno(9)
$$
According to the starting discussion in this section, we can describe the evolution
of the four velocity in frame $S'$ as
$$
\bar u'(\tau+\delta\tau)=A'(\tau) \bar u'(\tau), \eqno(10)
$$
where $A'(\tau)$ is the operator for an infinitesimal Lorentz boost in the $x'_2$
direction. From (4) and (5), $A'$ in matrix form is given by
$$
A'=\pmatrix{1&0&\delta v'_2&0\cr
	0&1&0&0\cr
	\delta v'_2&0&1&0\cr
	0&0&0&1\cr}. \eqno(11)
$$
Applying $A'$ to $\bar u'(\tau)$ given by (8) we get
$$
\bar u'(\tau+\delta\tau)=\pmatrix{u'^0+u'^2\delta v'_2\cr
	0\cr
	u'^0\delta v'_2+u'^2\cr
	0\cr}.\eqno(12)
$$
Now if we Lorentz transform $\bar u'(\tau)$ and $\bar u'(\tau+\delta\tau)$
to $S$, we find
$$
\bar u(\tau)=\pmatrix{\gamma u'^0\cr
		\gamma v u'^0\cr
		u'^2\cr
		0\cr},\eqno(13)
$$
and
$$
\bar u(\tau+\delta\tau)=\pmatrix{\gamma(u'^0+u'^2\delta v'_2)\cr
	\gamma v(u'^0+ u'^2\delta v'_2)\cr
	u'^2+u'^0\delta v'_2\cr
	0\cr},\eqno(14)
$$
where $\gamma=(1-v^2)^{-1/2}$ and the $4\times 4$ 
matrix for the {\it finite} 
Lorentz transformation to $S$ is
$$
\pmatrix{\gamma&\gamma v&0&0\cr
	\gamma v&\gamma&0&0\cr
	0&0&1&0\cr
	0&0&0&1\cr}.
$$
	
\bigskip
\bigskip

The question which now arises is: how should the change in the four velocity of the test
particle be regarded by an observer in $S$?
Since the $x_1$ component of $\bar u$ has also changed during $\delta\tau$,
it is clear that something more
involved than a single infinitesimal Lorentz boost along the $x_2$ axis
would be necessary in order to produce the observed change. The answer to this
question can be readily found in an intuitive way if we note that an accelerated
straight path in $S'$, along $x'_2$, would
be seen from the point of view of frame $S$ as an
accelerated but curved trajectory
in the $(x_1,x_2)$ plane which could be described as a succession
of infinitesimal boosts along the $x_2$ direction and rotations around $x_3$.
The operator performing this task is
$$
A=e^{-(\delta \phi_3 S_3- \delta v_2 K_2)}\simeq I-
{(\delta \phi_3 S_3-\delta v_2 K_2)} =\pmatrix{1&0&\delta v_2&0\cr
					0&1&\delta\phi_3&0\cr
					\delta v_2&-\delta\phi_3&1&0\cr
					0&0&0&0\cr}.\eqno(15)
$$
Solving  the linear system which comes out from $\bar u(\tau+\delta\tau)=A
\bar u(\tau)$, with $\bar u(\tau+\delta\tau)$ and $\bar u(\tau)$ given by
(13) and (14) respectively, and $A$ given by (15), we find
$$
	\delta v_2=\gamma\delta v'_2 \eqno(16)
$$
and
$$
	\delta\phi_3=\gamma v \delta v'_2.\eqno(17)
$$

Let us now write the familiar transformations equations between electric 
and magnetic fields in the two inertial frames $S$ and $S'$ (Jackson,
1975)
$$
E_1=E'_1, \
E_2=\gamma(E'_2+v B'_3), \
E_3=\gamma(E'_3-v B'_2).  
$$
$$
B_1=B'_1, \
B_2=\gamma(B'_2-v E'_3), \
B_3=\gamma(B'_3+v E'_2). \eqno(18)
$$

From the way in which we have set up our experiment, it is clear that
the rectilinear, accelerated, motion of the test particle
in $S'$, if charged, could be
attributed to a pure electric field in the $x'_2$
direction which would transform to $S$ according to
$$
E_2=\gamma E'_2,  
$$
$$
B_3=\gamma v E'_2.
$$

By comparison of these two equations with (16) and (17) we see that, at least
 for the particular case under consideration, an electric field in 
the $x'_2$ direction has the same transformation law that an 
infinitesimal Lorentz boost in that direction and that magnetic fields 
might, perhaps, have some relation with spatial rotations.

Let us now see see
how a pure spatial  rotation
of angle $\delta\phi'_3$ around the $x'_3$ axis performed on the four
velocity,
in $S'$, is seen by an
observer in $S$. In $S'$, the four velocity evolves according to
\bigskip

$$
\bar u'(\tau+\delta\tau)=\pmatrix{1&0&0&0\cr
				0&1&\delta\phi'_3&0\cr
				0&-\delta\phi'_3&1&0\cr
				0&0&0&1\cr} \pmatrix{u'^0\cr
	0\cr
	u'^2\cr
	0\cr}
=
\pmatrix{u'^0\cr
	u'^2\delta\phi'_3\cr
	u'^2\cr
	0\cr}.
\eqno(19)
$$
\bigskip

Following a similar procedure as the one used in the precedent case,
one finds, in $S$, that a spatial rotation around $x'_3$ of angle
$\delta\phi'_3 $ in $S'$, is
seen from the standpoint of an observer in $S$ as
a boost
$$
\delta v_2=\gamma v \delta\phi'_3 \eqno(20)
$$
in the $x_2$ direction followed by a rotation of angle
$$
\delta\phi_3=\gamma\delta\phi'_3 \eqno(21)
$$
around $x_3$. On the other hand, from
equations (18), it is readilly seen that a pure magnetic field along $x'_3$, in
$S'$, would transform to $S$ as
$$
E_2=\gamma v B'_3,
$$
$$
B_3=\gamma B'_3.
$$

We wish to show now that every relation between electric and magnetic 
fields in (18) has a geometric counterpart of whom equations (16), (17) 
and (20), (21) are particular cases. Following the same somewhat 
pedestrian, but clarifying approach, we consider first
a $x'_1$ boost in $S'$ acting on a generic four-velocity vector $\bar
u'(\tau)=(u'^0, u'^1, u'^2, u'^3)$, compute the change in $\bar u'(\tau)$,
Lorentz transform this change to $S$ and find the relation between the
operators responsible for the change in both inertial frames. As one 
should expect for longitudinal accelerations, the result is
$$
\delta v_1=\delta v'_1 \eqno(22)
$$
(as geometric counterpart of $E_1=E'_1$). Although somewhat laborious, 
it is also possible to show that no other combination of boosts and 
rotations in $S'$ yields a $\delta v_1 \neq 0$ in $S$. To prove that
$$
\delta\phi_1=\delta\phi'_1, \eqno(23)
$$
as counterpart of $B_1=B'_1$, it is only necessary to note that in a 
rotation of the spatial part of the four velocity around the preferred 
axis $x'_1$ only the components orthogonal to $u'_1$ are changed and 
these components remain without alteration under a Lorentz 
transformation in the $x_1$ direction.

In order to find geometric expressions similar to the second and last 
relations in (18) we shall consider a boost in the $x'_2$ direction 
followed by a rotation	around $x'_3$. The evolution equations are in $
S$
\bigskip

$$
\bar u'(\tau+\delta\tau)=\pmatrix{1 &0&\delta v'_2&0\cr
			0&1&\delta\phi'_3&0\cr
			\delta v'_2&-\delta\phi'_3&1&0\cr
			0&0&0&1\cr} 
			\pmatrix{u'^0\cr
			u'^1\cr
			u'^2\cr
			u'^3\cr}=
			\pmatrix{u'^0+u'^2\delta v_2 \cr
			u'^1+u'^2\delta\phi'_3\cr
			u'^0\delta v'_2-u'^1\delta\phi'_3+u'^2\cr
			u'^3\cr}. \eqno(24)
$$
\bigskip

On the other hand, the Lorentz transformed vectors to $S$ $\bar
u(\tau+\delta\tau)$ and $\bar u(\tau)$ are
$$
\bar u(\tau+\delta\tau)=\pmatrix{\gamma(u'^0+u'^2\delta v'^2)+\gamma v(
u'^1+u'^2\delta\phi'_3)\cr
	\gamma v(u'^0+u'^2\delta v'^2)+\gamma(u'^1+u'^2\delta\phi'_3 )\cr
	u'^0\delta v'_2-u'^1\delta\phi'_3+u'^2\cr
	u'^3\cr}, \eqno(25)
$$
and
$$
\bar u(\tau)=\pmatrix{\gamma(u'^0+v u'^1)\cr
		      \gamma(u'^1+v u'^0)\cr
		      u'^2\cr
		      u'^3\cr}.\eqno(26)
$$
The observed change in $S$ is thus
$$
\delta\bar u(\tau)=\pmatrix{\gamma (u'^2\delta v'_2+ v u'^2\delta\phi'
_3)\cr
	\gamma(v u'^2\delta v'_2+u'^2\delta\phi'_3)\cr
	u'^0\delta v'_2-u'^1\delta\phi'_3)\cr
	0\cr}.\eqno(27)
$$
We take now into account that, as we have 
previously seen, $\delta v'_1=0 \Rightarrow \delta v_1=0$ and $\delta\phi'
_1=0 \Rightarrow \delta\phi_1=0$. Also, since the $u'^3$ component have 
remained unchanged in both inertial frames: $\delta v_3=0,\ \delta\phi_2=
0$. The evolution equations (written now as $(A-I)\bar u(\tau)=
\delta\bar u(\tau)$) in $S$ are then
\bigskip

$$
\pmatrix{0&0&\delta v_2&0\cr
	0&0&\delta\phi_3&0\cr
	\delta v_2&-\delta\phi_3&0&0\cr
	0&0&0&0\cr}
	\pmatrix{\gamma(u'^0+v u'^1)\cr
		      \gamma(u'^1+v u'^0)\cr
		      u'^2\cr
		      u'^3\cr}
=\pmatrix{\gamma (u'^2\delta v'_2+ v u'^2\delta\phi'
_3)\cr
	\gamma(v u'^2\delta v'_2+u'^2\delta\phi'_3)\cr
	u'^0\delta v'_2-u'^1\delta\phi'_3)\cr
	0\cr},\eqno(28)
$$
\bigskip
{\noindent and solving for $\delta v_2$ and $\delta\phi_3$:}
$$
\delta v_2=\gamma(\delta v'_2+v \delta\phi'_3), \eqno(29)
$$
$$
\delta\phi_3=\gamma(v \delta v'_2+\delta\phi'_3). \eqno(30)
$$

In the same way, equations similar to the third and fifth of (18) can be 
obtained.

\bigskip
\bigskip

\noindent {\bf 4. Lorentz Force}

\bigskip

In the last two sections, we have gained some understanding of how  the
evolution of a particle's four velocity  can be related to infinitesimal boosts
and rotations in spacetime and with the way these geometric operations,
considered
here as linear operators, transform between inertial reference frames. We shall
now attempt to relate the Lorentz force to these 
infinitesimal transformations.
To this end, it is convenient to consider first  a
general infinitesimal Lorentz boost,
the explicit
form of $A(\tau)$ is
$$
A(\tau)=e^{(\delta\vec v.\vec K)} \simeq I+\delta\vec v.\vec
K=
$$
$$
=I+    \left( \matrix{0  &\delta v_1 &\delta v_2 &\delta v_3 \cr
	\delta v_1 &0 &0 &0 \cr
	\delta v_2 &0 &0 &0 \cr
	\delta v_3 &0 &0 &0 \cr} \right). \eqno(31)
$$

\bigskip

Expanding (3) with $A(\tau)$ given by (31), and $\bar u(\tau)=(u^0,\vec u)=
(\gamma, \gamma\vec v)$, we get
$$
u^0(\tau+\delta\tau)=u^0(\tau)+\delta\vec v.\vec u
$$
$$
u^i(\tau+\delta\tau)=u^i(\tau)+\delta v^i u^0=u^i(\tau)+\gamma
\delta v^i \eqno(32)
$$

$i=1, 2, 3.$
\bigskip

From  these equations, the change in the four
velocity $\delta\bar u$ is defined as
$$
\delta\bar u\equiv (\delta\vec v.\vec u, \gamma\delta\vec v).
\eqno(33)
$$
\bigskip

We note that $\delta\bar u$ is orthogonal to the four velocity
$\bar u=(\gamma, \gamma\vec v)$:
$$
\delta\bar u.\bar u=\gamma(\delta\vec v.\vec u)-\gamma^2\delta\vec
v.\vec v=0.
$$
\bigskip

Equations (32) can also be written as
$$
{u^0(\tau+\delta\tau)-u^0(\tau) \over\delta\tau}={\delta\vec v
\over\delta\tau}.\vec u,
$$
and
$$
{u^i(\tau+\delta\tau)-u^i(\tau) \over\delta\tau}={\delta
v^i\over\delta\tau} u^0 \eqno(34)
$$
$(i=1, 2, 3)$.
\bigskip

{\noindent If we define now}
$$
\chi_i(\tau)=\lim_{\delta\tau\to 0}{\delta v^i\over\delta\tau}, \eqno(35)
$$
we can write instead of (34):
$$
{du^0\over d\tau}=\vec\chi(\tau).\vec u
$$
and
$$
{du^i\over d\tau}=\chi_i(\tau) u^0. \eqno(36)
$$

\bigskip
For the case under consideration (i.e. infinitesimal Lorentz boosts), the last
equations are valid irrespective of the  external field.
(they do not describe the
more general kind of change as we have not included spatial rotations).
In the context of flat spacetime, we know that
any change in a particle's four velocity should arise from the
interaction with an external driving field. Due to this fact, we can
attempt to relate the quantities $\chi_i(\tau)$ with
the components of the electric
field.

To elucidate further this point let
us now examine the part of the Lorentz force that only depends
on the electric field:
$$
{du^0\over d\tau}={e\over m}\vec E.\vec u
$$
$$
{du^i\over d\tau}={e\over m}E_i u^0. \eqno(37)
$$
It is apparent that the last equations	are completely similar to
equations (36). Furthermore, experimentally we know that:

a) The electric force acting on a particle is proportional to its charge.

b) The electric force depends on the
point of spacetime where a charge is located and
not of $\vec u $.

It seems thus that  no conflict with previous notions
 about the electric field arises if we substitute in (36),
 up to a proportionality
constant, the quantities $\chi_i$ by the components of the electric field
 and consider this field as
responsible for the succession of infinitesimal Lorentz boosts that
change the four velocity of the particle.
(Note that we could write
now instead of (3)
$$
\bar u(\tau+\delta\tau)=e^{{e\over m}(\vec E.\vec
K)\delta\tau}\bar u(\tau) \eqno(38)
$$
and obtain the electric force).
\bigskip
To get the full electromagnetic force one should only
 notice that equations (36) cannot be completely general because
if the evolution in the four velocity vector can be described in terms of an
infinitesimal Lorentz boost in a certain frame, from the point of wiew of
another inertial frame we have seen in the preceding section that 
it is necessary, in general, to use the combination of
 Lorentz boosts and infinitesimal spatial rotations. As tensor equations
express physical laws in a form independent of the choice of inertial frame
one should expect to find the covariant form of the equation of motion
from a more general transformation on the four velocity.
As a matter of fact, although equations (36) are 
 compatible with the condition
${(du^\alpha / d\tau)}.u_\alpha =0$,
the
real limitation is that
it does not seem possible to write them
in the form of a tensor expression valid in any coordinate frame.
One is thus lead to conclude that equations  of the kind
of (36) and (37), should 
only be valid in a preferred frame of reference (see section 4
below).
By considering a more
general kind of motion, we shall now see that
 the  form of the Lorentz Force, and the number of non
null components of the field, is dictated by the specific form of the 6
matrices ($S_i$ and $K_i$) generators of the Lorentz Group

\bigskip

As previously noted, a general description of the 
evolution of the four velocity vector, valid in any frame of
reference, should  include
spatial rotations as well as boosts which means that
we need all the six generators of the Lorentz Group. The operator (31)
should be replaced
by

$$
	A=e^{-(\delta\vec\phi.\vec S-\delta\vec v.\vec K)}, \eqno(39)
	$$

We now relate the change in the velocity of the particle $\delta\vec v$
with an external force field which we call 
$\vec\epsilon(x)$ and the rotation $\delta\vec
\phi$ with another field $\vec b(x)$. The change in the velocity vector 
$\delta\vec v$	should be proportional to $\delta \tau$
and depend on the particle's spacetime location. Accordingly, we write

\bigskip

$$	\delta\vec v =k\vec\epsilon(x) \delta\tau, \eqno(40)
$$
\bigskip

\noindent {and, since from a geometric standpoint  a spatial rotation is 
quite similar to a space-time rotation (boost), we also write}
\bigskip

$$	\delta\vec\phi =k \vec b(x) \delta\tau, \eqno(41)
$$
$k$ being a  constant.
\bigskip

\noindent {Since $A$ is infinitesimal:}
\bigskip

$$A(x) \simeq I - k[\vec b(x).\vec S-\vec\epsilon(x).\vec K ]
\delta\tau. \eqno(42)
$$

\bigskip

\noindent {Once expanded with $A(x)$ given by (42), equation (3) 
 now reads}
\bigskip

$$ \left( \matrix{u^0(\tau+\delta\tau) \cr
	u^1(\tau+\delta\tau)	       \cr
	u^2(\tau+\delta\tau)	       \cr
	u^3(\tau+\delta\tau)	       \cr} \right) \simeq
	\left( \matrix{u^0(\tau)       \cr
	u^1(\tau)			\cr
	u^2(\tau)			\cr
	u^3(\tau)			\cr} \right)+ k\delta\tau
   \left( \matrix{0  &\epsilon_1 &\epsilon_2 &\epsilon_3 \cr
	\epsilon_1 &0 &b_3 &-b_2 \cr
	\epsilon_2 &-b_3 &0 &b_1 \cr
	\epsilon_3 &b_2 &-b_1 &0 \cr} \right).
	\left( \matrix{u^0(\tau) \cr
		u^1(\tau) \cr
		u^2(\tau) \cr
		u^3(\tau) \cr} \right). \eqno(43) $$
\bigskip

\noindent {In the limit $\delta\tau \rightarrow 0$ the former matrix equation
can be expressed as the following two equations}

\bigskip

$$  {du^0\over d\tau}=k\vec u.\vec\epsilon \eqno(44)
$$
\bigskip

\noindent {and}
\bigskip
 $$ {d\vec u \over d\tau}=k(\vec\epsilon u^0 + \vec u 
 \times \vec b), \eqno(45)
    $$
\bigskip

\noindent {which for $k\equiv e/m$ are identical to the equations 
    describing the Lorentz force.}

From (43) and recalling that contravariant vectors are represented by
column matrices, the last two equations can be joined into a single tensor
expression:
$$
{du^\alpha \over d\tau}=kF^\alpha_\beta u^\beta, \eqno(46)
$$
where  $F^\alpha_\beta$ is associated with the {\it antisymmetric}
electromagnetic tensor by
$$
F^\alpha_\beta= \eta^{\alpha\delta}F_{\delta\beta}, \eqno(47)
$$
$\eta^{\alpha\delta}$  being the metric tensor.

\bigskip

\bigskip

\noindent {5. \bf Final Remarks}
\bigskip

A very special situation  which deserves a comment is the situation
corresponding to a charged test particle moving in circular motion
around a charged object. As the energy remain constant in this
case, an observer in the rest frame of the source might conclude that the test
particle is performing a succession of rotations. However notice
that the equations of
motion in this case are, from (44) and (45)
$$
{du^0\over d\tau}=k\vec u.\vec\epsilon=0  \eqno(48)
$$
and
$$
{d\vec u \over d\tau}=k\vec\epsilon u^0.  \eqno(49)
$$
From (43), it
is readily seen that the last equations are related to boosts not
to rotations (The reader can convince himself that an
infinitesimal boost orthogonal to $\vec u$ is both compatible, to
first order, with the normalization condition $u^\alpha
u_\alpha=1$ and with the conservation of $|\vec u|$).
 This kind of motion, as well as any other taking
place under a central electrostatic field, should
be understood, as a sort of
{\it continuous fall}, in terms of infinitesimal boosts when described in the
preferred frame in which the source is at rest. The
opposite situation would be a particle undergoing circular or
helicoidal motion in a constant magnetic field which should be
understood in terms of infinitesimal rotations.

Finally note that the interpretation of magnetic fields as performers of 
infinitesimal spatial rotations give also a geometric explanation of why 
there is no magnetic force in the rest system of a charged particle
and in the direction of the field.
Being the spatial components of the four velocity null, there is nothing
to rotate and, also, the longitudinal component of the
velocity remain unchanged in a rotation.
However, there is no way of avoiding the existence of
magnetic fields as it is impossible, for two or more source charges in 
relative motion, to set up a preferred reference system in which all the 
sources are at rest.

\bigskip

\vfill
\eject

{\noindent  \bf References}

{\noindent  Barut A O 1964 {\it Electrodynamics and Classical Theory
of Fields and Particles}  (Dover Publications Inc.), Chapter 2.}

{\noindent  French A P 1968 {\it Special Relativity}  (Norton,
New York), Chapter 8}

{\noindent   Kobe D H 1986 Am. J. Phys. {\bf 54} (7), 631-636 }

{\noindent  Lorrain P and  Corson D R 1970 {\it Electromagnetic Fields and
Waves},  2nd edition (W.H. Freeman, San Francisco), Chapter 6.}

{\noindent   Jackson J D 1975 {\it Classical Electrodynamics}
(Wiley, New York), Chapter 11 and page 578.}

{\noindent  Rindler W 1982 {\it
Introduction to Special Relativity}  (Oxford
University Press), Chapter 6}.

{

\vfill
\eject

\end